\documentclass[12pt,a4paper]{article}

\usepackage{cite}
\usepackage{amsmath}
\usepackage{amssymb}
\usepackage{tikz}
\usepackage{fixltx2e}
\usepackage{amsfonts}
\usepackage{graphicx}
\usepackage{subfig}
\usepackage{caption}
\usepackage{dcolumn}
\usepackage{bm}
\usepackage[T1]{fontenc}
\usepackage{textcomp}

\usepackage[
colorlinks,linkcolor=blue,citecolor=magenta]{hyperref}
\usepackage[top=3cm,right=2cm,bottom=2.5cm,left=2cm]{geometry}
\begin{document}
\title{Hamiltonian structure of bi-gravity, problem of ghost \\ and bifurcation}

\author{Zahra Molaee$^{a, b}$\thanks{zmolaee@ipm.ir}, Ahmad Shirzad$^{a,b}$\thanks{shirzad@ipm.ir}\\
	{\it $^{a}$Department of Physics, Isfahan University of Technology, Isfahan ,Iran}\\
	{\it$^{b}$School of Particles and Accelerators,}\\
	{\it Institute for Research in Fundamental Sciences (IPM),}\\
	{\it P.O.Box 19395-5531, Tehran, Iran}}
\maketitle
\begin{abstract}
We analyze the Hamiltonian structure of a general theory of bi-gravity where the interaction term is a scalar function of the form $ V(\mathcal{X}^{n}) $  where $ \mathcal{X} $ may be $\sqrt{g^{-1}f}  $ or $ g^{-1}f $. We give necessary conditions for the interaction term of such a theory to be  ghost free.  We give a precise constraint analysis of the bi-gravity theory of Hassan- Rosen and show that the additional constraint which omit the ghost is our choice at the bifurcation point.
\end{abstract}
\section{Introduction}
Finding a consistent covariant theory of massive gravity is an old dream for about eight decades, beginning by the pioneer paper of Fierz and Pauli \cite{wit1}, in 1939. The main difficulty is arising ghosts in the spectrum of solutions. In recent years, there was made some hopes toward a consistent theory of massive gravity due to dRGT model \cite{drgt1}. Then Hassan and Rosen improved the model \cite{mmg} by replacing the flat background metric with an external metric $ f_{\mu\nu}$.  The interaction term added to Hilbert-Einstein action in this model is a  polynomial of the function $ Tr\sqrt{g^{-1}f}$. In order to have a covariant model, they introduced their bi-gravity model afterwards by giving dynamics to the second metric, via introducing the kinetic term $ \sqrt{-^{(4)}f}\mathcal{R}(f) $ in the Lagrangian. 

Bi-gravity model by itself is attractive theoretically as well as observationally in describing physical events. 
For instance, it has been recently shown \cite{YA} that doubly coupled models of bi-gravity are tightly constrained by observation in light of the neutron star merger GW170817/GRB170817A \cite{ligo}. These constraints indicate that viable bi-gravity theories would be singly-coupled, in that matter couples to only one of the two available metrics. Our focus here is on theoretically consistent bi-gravity models, specifically those that are ghost-free. Such ghost-free models should enable us to adjust the corresponding couplings to matter in a physically viable manner.

To investigate the existence of a ghost (or ghosts) a popular way is to expand the metric (or the metrics in bi-gravity) around a given background and search for conditions of avoiding negative kinetic terms. However, this method is not trusty enough, since  just acquires information in the vicinity of the given background solution. The next method, which is much more trustworthy, is the Hamiltonian analysis of the dynamical structure of the model. This approach, however, is much more complicated and requires lengthy and tedious calculations.  

For the massive gravity, the Hamiltonian analysis given in \cite{HR1} shows that ghost disappears. Despite of some doubts in Refs. \cite{klu1}-\cite{klu3} we show in our previous paper \cite{MS} that in full phase space of 20 variables, there is no ghost in massive gravity. Concerning the case of HR bi-gravity, a crucial calculation is done by Hassan and Rosen \cite{HR5} to show that additional constraints emerge in the Hamiltonian analysis of the theory which lead to omitting the ghost degrees of freedom. Based on this observation, the new model of HR bi-gravity gained considerable attraction among the community. Hence, the Hamiltonian analysis of HR bi-gravity, for assuring people about additional constraints, is a very important task which may validate or invalidate hundreds  of papers based on reliability of calculations of a few papers  written on this issue \cite{HR5}-\cite{solv2}. 

However, we think that deducing additional constraints needed to omit the Boulware-Deser ghost does not come true completely. In other words,   
 the main reference on this issue, i.e. Ref \cite{HR5}, contains subtleties which contradict the standard Dirac approach for constrained systems. In fact, the additional constraint $ \mathcal{C}_{2} $ which has the crucial role of omitting the ghost is just the Poisson bracket  $ \{\mathcal{C},\mathcal{D}\} $ of two existing constraints $ \mathcal{C} $ and $\mathcal{D}$. In the context of Dirac formalism when  $ \{\mathcal{C},D\}\neq 0 $, it turns out that they are second class, while in the mentioned  papers the constraint  $\mathcal{D} $ is considered as  a first class constraint on the basis of demanding enough number of first class constraints to generate diffeomorphism. Hence, it seems that the additional constraints needed to omit the ghost do not emerge naturally in the constraint structure of the model.
 
 Our main interest in this paper is to investigate more deeply the constraint structure of the bi-gravity and see how additional constraints may emerge to cancel the Boulware-Deser ghost.  As we will show, the crucial point is that the dynamical behavior of a system, including the number of degrees of freedom and the symmetries, may be different in some subregion of the phase space. For example, the problem of ghost may be solved only in some special subregion of the phase space. This may happen due to  the problem of bifurcation. Whenever we find multiplicative constraints, the theory may bifurcate into different branches with distinct physical properties. Our final answer to the problem of ghost in bi-gravity is that the theory is ghost free in one branch at the bifurcation point.
 
A second reason to study the constraint structure of the bi-gravity theory is that the original papers on the canonical analysis of HR bi-gravity has performed calculations in a 24 dimensional phase space containing $ g_{ij} $, $ f_{ij} $ (i.e. the spatial part of the metrics) and their conjugate momenta. In this approach, the lapse and shift functions have been considered as auxiliary fields. However, we think that a Hamiltonian analysis in the 40 dimensional phase space including lapse and shift functions as dynamical variables is more fundamental, since they are parts of metrics which do participate in dynamics as well as the gauge symmetry (i.e. diffeomorphisms) of the theory.  In fact, in the Hamiltonian formulation the momenta conjugate to the lapse and shift functions should play some roles in generating the gauge transformations.
 
Although the author of Ref. \cite{jklu} have also  tried to give a careful Hamiltonian analysis in 40 dimensional phase space, he finally found two similar  differential equations for the lapse functions as the result of consistency of the constraints $ \mathcal{C}$ and $\mathcal{D}$. In his approach,  no additional constraint is  obtained to omit the ghost. On the other hand, there are not enough first class constraints for generating the full space-time diffeomorphism. The same author analyzed, in another paper \cite{Kluson}, a bi-gravity theory in which the interaction term is a function of $ Tr(g^{-1}f) $,  (rather than $ Tr\sqrt{g^{-1}f} $). He concluded finally that it is highly improbable to find a ghost free bi-gravity which supports the diagonal diffeomorphism as well. 

 Our Hamiltonian analysis in this paper is not limited to HR bi-gravity; we try to give a compelling Hamiltonian analysis for a general bi-gravity model with an interaction term $V$ as a polynomial function of  either $ Tr\sqrt{(g^{-1}f)^{n}}  $ or $ Tr((g^{-1}f)^{n}) $. We show that, in every parametrization of the lapse and shift functions, the most determinant factor for the presence of ghosts is the matrix of second derivatives of $ V $ with respect to lapses and shifts. As we will see, one needs, as a necessary condition, four null-vectors for this matrix to guarantee the diffeomorphism gauge symmetry and one more null-vector for omitting the ghost.  
 
  In section 2, we give a general framework for the Hamiltonian analysis of bi-gravity models, and the crucial role of the second derivatives of the interaction potential with respect to lapses and shifts. In section 3, we give our main Hamiltonian analysis of the HR gravity. In section 4, we analyze a model without  square root, using a different set of lapse and shift variables. We show that it is not improbable to have a ghost free model of this kind. Section 5, denotes some concluding remarks and some view points towards future works.
  
\section{ Hamiltonian structure of general bi-gravity}

We present a general framework for analyzing a bi-gravity theory. Consider a dynamical theory in four dimensions with two spin-2 fields $ f_{\mu\nu} $ and $ g_{\mu\nu} $
described by the following action 
\begin{equation}
S=\int d^{4}x \left(  M_{g}^{2}\sqrt{-^{(4)}g} 
\mathcal{R}(g)+ M_{f}^{2}\sqrt{-^{(4)}f} 
\mathcal{R}(f)+2m^{4} \sqrt{-^{(4)}g} V(\mathcal{Z}^{\mu}_{\ \nu})\right) , \label{m11}
\end{equation}
where  $  \mathcal{Z}^{\mu}_{\ \nu}=g^{\mu\rho}f_{\rho \nu}$, $M_{g}$ and $ M_{f}$ are Plank masses and $  m $ is mass parameter. 
Note that  $g^{\mu\nu}  $ is the inverse of $g_{\mu\nu}  $ , while we do not use $ f^{\mu\nu} $ as the inverse of $ f_{\mu\nu} $ except in construction of the curvature $ \mathcal{R}(f) $ . The interaction potential  $ V(\mathcal{Z}^{\mu}_{\ \nu})$ is a scalar function of the matrix $ \mathcal{Z} $. This can include $ Tr(\mathcal{Z}) $ or more generally $Tr(\mathcal{Z}^{n})$.
In ADM formalism, the metrics has the following (3+1) decomposition \cite{berg}, 
\begin{equation}
 g_{\mu\nu}=
\left( \begin{array}{cr}
-N^{2}+N_{i}N^{i} & N_{i}  \\
N_{i} &  g_{ij} \\
\end{array}\right),
 f_{\mu\nu}=
\left(\begin{array}{cr}
-M^{2}+M_{i}M^{i}       & M_{i}   \\
M_{i}  & f_{ij}  \\   \label{g1}  
\end{array}\right)
\end{equation}
where $N,M,N^{i},M^{i}$ are called lapses and shifts respectively.
The inverse  metrics $g^{\mu\nu}  $ and  $f^{\mu\nu} $ can be written  as
\begin{equation}
g^{\mu\nu}=
\left( \begin{array}{cr}
-N^{-2}       & N^{i}N^{-2}  \\   
N^{i}N^{-2}      & g^{ij}-N^{i}N^{j}N^{-2}  \\
\end{array}\right),\ \ \ \ 
f^{\mu\nu}=
\left(\begin{array}{cr}
-M^{-2}   & M^{i}M^{-2}  \\   
M^{i}M^{-2}  & f^{ij}-M^{i}M^{j}M^{-2}  \label{g2}  
\end{array}\right).
\end{equation}
Note that in the interaction term we do not need to raise the indices of $f_{\mu\nu}$, while the indices in g-sector will raise and lower with $g^{\mu\nu}$ and $g_{\mu\nu}$. Since the interaction term does not depend on the derivatives of the fields, the momenta $ \Pi_{N},\Pi_{N^{i}},\Pi_{M} $  and $ \Pi_{M^{i}}$ are primary constraints and the Lagrangian density reads  
\begin{equation}
\mathcal{L}=\pi^{ij}\dot{g_{ij}}+p^{ij}\dot{f_{ij}}-H_{c},\label{g3}
\end{equation}
where $ \pi^{ij} $ and $p^{ij}  $ are conjugate momenta of $ g_{ij} $  and $ f_{ij} $  respectively and 
\begin{equation}
H_{c}=\int d^{3}x \left(N^{\mu} \mathcal{R}^{(g)}_{\mu}+M^{\mu} \mathcal{R}^{(f)}_{\mu}+\mathcal{V}\right) . \label{m12}
\end{equation}
The expressions $\mathcal{R}^{(g)}_{0}$ , $\mathcal{R}^{(g)}_{i}$   are the Hamiltonian and momentum constraints of the corresponding  Hilbert-Einstein action of the metric $g_{\mu\nu}$ as follows
\begin{equation}
\mathcal{R}_{0}^{(g)}= M_{g}^{2}\sqrt{g}\mathcal{R}+\dfrac{1}{ M_{g}^{2}\sqrt{g}}(\frac{1}{2}\pi^{2}-\pi^{ij}\pi_{ij}),\hspace{10mm} \mathcal{R}_{i}^{(g)}=2\sqrt{g}g_{ij} \triangledown _{k}(\frac{\pi^{jk}}{\sqrt{g}}).\label{a14}
\end{equation}
Similar relations should also be considered for $\mathcal{R}^{(f)}_{0}$ and $\mathcal{R}^{(f)}_{i}$ in terms of the $f$-metric. Noticing that $ \sqrt{-^{(4)}g} =N\sqrt{g} $, where $ g\equiv det(g_{ij}) $,
 the interaction term reads 
\begin{equation}
\mathcal{V}= 2m^{4} N \sqrt{g} V(\mathcal{Z}^{\mu}_{\nu}).\end{equation}
Let us denote the whole set of lapse and shift functions as 
$ L^{a},\ a=1,...,8 $ where the first four  refer to $ N $ and $ N_{i} $ and the remaining ones to $ M $ and $ M_{i} $. In this way the canonical Hamiltonian (\ref{m12}) reads
\begin{equation}
H_{c}=\int d^{3}x \left( L^{a}\mathcal{R}_{a}+\mathcal{V}\right) , \label{m16}
\end{equation}
 where the same notation has been used to denote $ \mathcal{R}^{(g)}_{0}, \mathcal{R}^{(g)}_{i}, \mathcal{R}^{(f)}_{0} $ and $\mathcal{R}^{(f)}_{i}$ as 
 $ \mathcal{R}_{1},... , \mathcal{R}_{8} $.
    The total Hamiltonian reads 
\begin{equation}
H_{T}=H_{c}+\int d^{3}x u^{a}\Pi_{a}, \label{m13}
\end{equation}
where $\Pi_{a}$, as primary constraints, are momenta   conjugate to $L^{a}$  and $u^{a}$ are Lagrange multipliers. 
The  primary constraints should be preserved during the time. This gives the second level constraints as
\begin{equation}
\mathcal{\mathcal{A}}_{a}\equiv \lbrace \Pi_{a}, H_{c} \rbrace=-(\mathcal{R}_{a}+\frac{\partial \mathcal{V}}{\partial L^{a}})\approx 0.\label{m14}
\end{equation}
The constraints $\mathcal{\mathcal{A}}_{a}$ should also be preserved during the time, i.e.
\begin{equation}
 \lbrace \mathcal{A}_{a}, H_{T} \rbrace= \lbrace \mathcal{A}_{a}, H_{c} \rbrace-  \frac{\partial^{2} \mathcal{V}}{\partial L^{a}\partial L^{b}} u^{b}  \approx 0.\label{m15}
\end{equation}
 
We know that the bi-gravity theory is diffeomorphic invariant. Hence, loosely speaking, we  demand that four arbitrary fields exist in the dynamical analysis of the theory. This can be achieved by demanding that at least four Lagrange multipliers $ u^{a} $ should remain undetermined. In other words, the rank of the matrix $ \partial^{2} \mathcal{V}/\partial L^{a}\partial L^{b} $ should not exceed four, in order  to have at least four null-vectors. If there were no interaction, we would have eight null-vectors due to vanishing  the matrix $ \partial^{2} \mathcal{V}/\partial L^{a}\partial L^{b} $. Suppose $ \chi^{a}_{(\alpha)} $ are null-vectors of  $ \partial^{2} \mathcal{V}/\partial L^{a}\partial L^{b} $. Then from Eq. (\ref{m15}) we find the following third level constraints  $ \mathcal{B} _{(\alpha)} $ labeled by the index $ \alpha $,

\begin{equation}
\mathcal{B} _{(\alpha)} \equiv \chi^{a}_{(\alpha)} \{\mathcal{A}_{a},H_{c}\} \approx 0,\label{Vc18}
\end{equation}
However, some of $ \mathcal{B} _{(\alpha)} $'s may vanish on the constraint surface. For the case of no interaction, we have two disjoint Einestain-Hilbert theories and the expressions  $ \mathcal{B} _{(\alpha)} $  consist of Poisson brackets of $ \mathcal{R}^{(g)}_{0}, \mathcal{R}^{(g)}_{i}, \mathcal{R}^{(f)}_{0} $ and $\mathcal{R}^{(f)}_{i}$ which vanish weakly.  For a generic interaction, we also expect that at least four of the third level constraints $\mathcal{B} _{(\alpha)}   $ are trivial  due to our need to have at least four secondary first class constraints to generate diffeomorphisms. If more than four $\mathcal{B} _{(\alpha)} $ vanish, we would have extra symmetries besides diffeomorphisms and  the theory would have less number of degrees of freedom, comparing to what we consider in the following. On the other hand, consistency of the third level constraints (if any) should not determine the Lagrangian multipliers. Therefore, it is legitimate to assume that at least four of the expressions $\mathcal{B} _{(\alpha)} $  should vanish weakly. We will discuss this point with more details for two distinct examples in the following sections. 

To be used in the next section, let us consider the possibility of redefinition of the second level constraints. In the framework of constrained systems, one may replace, for  some reasons, the constraints $ \mathcal{A}_{a} $  with $  \mathcal{\tilde{A}}_{a}  $ such that  
\begin{equation}
\mathcal{A}_{a}\approx0 \Leftrightarrow \mathcal{\tilde{A}}_{a} \approx0. \label{jj}
\end{equation}
 Hence, the equation (\ref{m15}) should be replace by
 \begin{equation}
 \{\mathcal{\tilde{A}}_{a}, H_{T}\}= \{\tilde{\mathcal{A}}_{a}, H_{c}\}+\frac{\partial \mathcal{\tilde{A}}_{a} }{\partial L^{b}}u^{b}. \label{jj1}
 \end{equation}
 In this way, our discussions after Eq. (\ref{m15}) are valid by considering the null-vectors of the matrix $ \partial  \mathcal{\tilde{A}}_{a}/\partial L^{b}$ instead of $\partial^{2} \mathcal{V}/\partial L^{a}\partial L^{b} $.
 
If the rank of  $\partial  \mathcal{\tilde{A}}_{a}/\partial L^{a}$ is four, and we have no further third level constraint $\mathcal{B} _{(\alpha)} $,  this means that four of the lapse-shift functions $L^{\bar{a}} $ and the corresponding second level constraint $ \tilde{A}_{\bar{a}}$ are first class and the remaining $ L^{\tilde{a}} $ as well as $ \tilde{A}_{\tilde{a}}$ should be second class. 

Remember the famous formula of the number of phase space degrees of freedom in a constrained system reads 
\cite{zms2}
\begin{equation}
DOF= \mathcal{N}-2FC-SC,\label{hy}
\end{equation}
where $  \mathcal{N} $ is the number of original variables, $ FC $ is the number of first class constraints and $ SC $ is the number of second class constraints. For the current case of 40 phase space variables with 8 first class and 8 second class constraints we find
\begin{equation}
DOF=40-2 \times 8 - 8=16,
\end{equation}
which correspond to 8 degrees of freedom in the configuration space. This can be interpreted as one  massive and one massless gravitons accompanying by a scalar ghost field. 

In order to omit the ghost degree of freedom, we need to find at least two more second class, or one more first class constraints. The latter possibility corresponds to one more gauge symmetry besides diffeomorphism, which does not sound well. Moreover, a first class constraint in the second level implies one more primary first class constraint. Hence, it is not reasonable to have only one more first class constraint. So, in order to omit the ghost we should expect to find two more second class constraints. 

To reach this goal we need a fifth null-vector for the matrix  $\partial  \mathcal{\tilde{A}}_{a}/\partial L^{a}$   which leads to a new constraint at the third level via Eq. (\ref{Vc18}), i.e. 
 \begin{equation}
 \mathcal{B} \equiv \chi^{a}_{(5)} \{\mathcal{A}_{a},H_{c}\} \approx 0.\label{Vc181}
 \end{equation}
 If the new constraint depends on lapse-shift functions, one combination of the Lagrange multipliers $ u^{a} $ in Eq. (\ref{m13}) would be determined as the result of consistency of the constraint $ \mathcal{B} $.\footnote{There is a technical point here, i.e.  three variables $L^{\tilde{a}}$ (see before Eq.(\ref{hy})) are determined in the second level of consistency in terms of the canonical variables. Moreover, the corresponding momenta $\Pi_{\tilde{a}}$ and second level constraints $ \tilde{\mathcal{A}}_{\tilde{a}} $ constitute a system of second class constraints. 
 	Therefore, for the next level of consistency one should consider the Dirac brackets instead of Poisson brackets. This implies that the constraint equations $\Pi_{\tilde{a}} \approx 0$ and  $ \tilde{\mathcal{A}}_{\tilde{a}}\approx 0 $ should be imposed as strong equalities, i.e.  $\Pi_{\tilde{a}}= 0$ and  $ \tilde{\mathcal{A}}_{\tilde{a}}= 0 $ . Hence, the terms $u_{\tilde{a}} \Pi_{\tilde{a}}  $ in the total Hamiltonian disappears at all. So when we say that consistency of the third level constraints may determine Lagrange multipliers, we mean the remaining ones other than  $ u_{\tilde{a}} $'s. } Hence, the constraint analysis would stop here with just one more second class constraint. This leads to a phase space with 15 dynamical fields. This may sounds undesirable to have a phase space with odd number of dynamical degrees of freedom. However, as shown in \cite{comeli2} and \cite{Henu} this does not mean an odd-dimensional phase space for field  theories. Meanwhile, the main problem is we need one more second class constraint to omit the ghost. 
 
 Let us summaries the final conclusion of this section. In order to have a ghost free bi-gravity theory, we need to have a diffeomorphic invariant interaction with two following properties.
 
 i) The rank of the matrix $\partial^{2} \mathcal{V}/\partial L^{a}\partial L^{b} $ or  $\partial  \mathcal{\tilde{A}}_{a}/\partial L^{a}$, in the case of redefinition of the constraints, should be three.
 
 ii) The new constraint $ \mathcal{B} $ emerged due to the fifth null-vector should not contain lapse-shift functions.
 
 We will investigate the above conditions in two approaches given  in the following sections. 
 
\section{Hamiltonian analysis of HR Bi-gravity }
We start by investigating the Hamiltonian formulation of HR bi-gravity  given by the following action\cite{mmg2},
\begin{equation}
S=M^{2}_{g} \int d^{4}x \sqrt{-g} 
R(g)+M^{2}_{f}\int d^{4}x \sqrt{-f} 
R(f)+2m^{4} \int d^{4}x \sqrt{-g}  \sum_{n=0}^{4} \beta_{n}e_{n}(\Bbbk). \label{a1}
\end{equation}
In Eq. (\ref{a1}) $\beta_{n}$ are free parameters, $m$ is a mass parameter, $M_{g}$ and $ M_{f}$ are Plank masses  and  $\Bbbk \equiv \sqrt{g^{-1}f}$ where $(g^{-1}f) ^{\mu}_{\ \nu}= g^{\mu\lambda}f_{\lambda\nu}$. The elementary symmetric polynomials $e_{n}(\Bbbk)$ are given in the appendix A.
In this paper we consider only minimal model of the interaction term where the coefficients $\beta_{n}$ are $ \beta_{0}=3, \  \beta_{1}=-1,\  \beta_{2}=\beta_{3}=0, \  \beta_{4}=1. $
By applying the following redefinition for the shift functions \cite{mmg2}
\begin{equation}
N^{i}=M n^{i}+M^{i}+N D_{\ j}^{i}n^{j},\label{a15}
\end{equation}
and choosing the $3\times3$ matrix $D^{i}_{\ j}$ appropriately (see appendix A), the interaction term as well as the whole action would become linear in the lapses  $N$ and $ M$ and shifts  $M^{i}$. Since the interaction does not involve derivatives of the metrics, the definitions of the momentum fields are similar to Hilbert-Einstein action as
\begin{eqnarray}&&
\pi^{ij}=-\sqrt{g}(K^{ij}-g^{ij}K)\label{c1},\\&&
p^{ij}=-\sqrt{f}(L^{ij}-f^{ij}L),\\&&
P_{M_{i}}\approx 0, P_{M}\approx 0,P_{N} \approx 0,P_{n^{i}}\approx 0,\label{bi9}
\end{eqnarray}
where $K^{ij}$ and $L^{ij}$  are three dimensional extrinsic curvatures due to $g$ and $f$ metrics respectively. As is seen, from Eq. (\ref{bi9})   we have 8 primary constraints $P_{M}$, $P_{N}$, $P_{M_{i}}$ and $P_{n^{i}}$. The Lagrangian density reads
\begin{eqnarray}&&
\mathcal{L}=M^{2}_{g}\pi^{ij}\partial_{t}g_{ij}+M^{2}_{f}p^{ij}\partial_{t}f_{ij} -H_{c}, \label{bi4}
\end{eqnarray}
with the canonical Hamiltonian  
\begin{equation}
H_{c}=M^{i} \mathcal{R}_{i} + M \mathcal{D} + N \mathcal{C},\label{bi5}
\end{equation}
in which
\begin{eqnarray}&&
\mathcal{C}=M^{2}_{g}\mathcal{R}_{0}^{g}+M^{2}_{g} D^{i}_{\ k} n^{k} \mathcal{R}_{i}^{g}-2m^{4}(\sqrt{g}\sqrt{x} D_{\ k}^{ k}-3\sqrt{g}),\\ &&\label{bi57}
\mathcal{D}=M^{2}_{f}\mathcal{R}_{0}^{f}+ M^{2}_{g} n^{i}\mathcal{R}_{i}^{g}-2m^{4}(\sqrt{g}\sqrt{x}-\sqrt{f}),\\ && \label{bi7}
\mathcal{R}_{i}= M^{2}_{g} \mathcal{R}_{i}^{g}+M^{2}_{f}\mathcal{R}_{i}^{f},
\label{bi6}
\end{eqnarray}
where $ x=1-n^{i}f_{ij}n^{j}$. 
 As a special case of equation (\ref{m13}) the total Hamiltonian reads  
\begin{eqnarray} &&
\mathcal{H}_{T}=\mathcal{H}_{c}+uP_{N}+vP_{M}+u^{i}P_{M^{i}}+v^{i}P_{n^{i}}, \label{k111}
\end{eqnarray} 
where $ u,v,u_{i} $ and $v_{i}$ are 8 undetermined Lagrange multipliers (8 fields, in fact). Since $N$, $M$ and $M_{i}$ appear linearly  in the canonical Hamiltonian,
consistency of the primary constraints $P_{M}$, $P_{N}$, $P_{M_{i}}$  (using the fundamental Poisson bracket given in appendix A) gives 5 secondary constraints as follows
\begin{eqnarray} &&
\lbrace P_{N},\mathcal{H}_{T} \rbrace= -\mathcal{C}\approx 0,
\\&& \lbrace P_{M},\mathcal{H}_{T} \rbrace=-\mathcal{D}\approx 0,
\\&& \lbrace  P_{M^{i}},\mathcal{H}_{T} \rbrace=-\mathcal{R}_{i}\approx 0.
\end{eqnarray}
However, all the terms of the canonical Hamiltonian involve the variables $n_{i}$. Hence, for consistency of $P_{n^{i}}$, we find directly 
\begin{eqnarray} &&
\lbrace P_{n^{i}},H_{c}\rbrace \equiv -\mathcal{S}_{i} =
-\left(M \delta^{k}_{\ i}+N \frac{\partial(D^{k}_{\ j}n^{j})}{\partial n^{i}}\right)U_{k} \approx 0,\label{bi13}
\end{eqnarray}
where
\begin{equation}
U_{k}=M^{2}_{g}\mathcal{R}_{k}(g)-2m^{4}\sqrt{g}n^{l}f_{lj}\delta^{j}_{\ k}x^{-1/2} \approx 0.\label{bi15}
\end{equation}
In this  way, for the current model, the secondary constraints $  \mathcal{A}_{a} $ of the previous section are $ -\mathcal{C},  -\mathcal{D},  -\mathcal{R}_{i}$ and $ -\mathcal{S}_{i} $ respectively. The matrix within the parenthesis on the right hand side of Eq. (\ref{bi13}) is the Jacobian of the transformation given in Eq. (\ref{a15}) which is invertible. Hence, Eq. (\ref{bi13})  leads to secondary constraint $U_{k}\approx 0$. So, we replace the secondary constraints with the new set $\mathcal{C},\mathcal{D},\mathcal{R}_{i}$ and $U_{i}$. This replacement is important. Notice that one may consider subregions of the phase space where the matrix  $(M \delta^{k}_{\ i}+N \frac{\partial(D^{k}_{\ j}n^{j})}{\partial n^{i}})$ is not full rank. This implies constraints which depend on the lapses $N$ and $M$, which would be second class with respect to the primary constraints  $P_{N}$ and $ P_{M} $. Here, we choose to put  away such possibilities.

Considering the problem from the opposite side, when the consistency condition leads to equality $\mathcal{S}_{i}= K_{ij}U_{j}\approx 0 $, we have two possibilities, either assume $ |K|\neq 0 $ which implies  $U_{j}\approx 0 $ instead of $\mathcal{S}_{i}\approx 0$; or assume that $ |K|\approx 0 $ and $U_{j}\neq 0 $ for some nontrivial null-vectors of $ K $. From this point of view the emerging constraints  $\mathcal{S}_{i}$ exhibit a bifurcation problem, and we should decide which way to follow in the rest of the problem. However,  we are not allowed to keep the original form of the constraints  $\mathcal{S}_{i}$, since this means we are mixing the two distinct possibilities simultaneously.

In our case, considering the constraints  $\mathcal{S}_{i}$ as given in Eq. (\ref{bi13}) leads to non vanishing Poisson brackets $ \{\mathcal{S}_{i},P_{M}\} $ and  $ \{\mathcal{S}_{i},P_{N}\} $. Hence, when we say that the matrix $ \left(M \delta^{k}_{\ i}+N \frac{\partial(D^{k}_{\ j}n^{j})}{\partial n^{i}}\right) $ is invertible, we mean, in fact, that we choose to leave in regions of phase space where this matrix is non-singular.
 \footnote{If the matrix elements $ K_{ij} $ where constant numbers, there were no difference in employing $ \mathcal{S}_{i} $'s or $ U_{i} $'s as constraints, since $ U_{i} $ where just linear combinations of $ \mathcal{S}_{i} $. However, if $ K_{ij} $'s depend on phase space variables (as in our case), then there is difference in choosing the set $\mathcal{S}_{i} $ or $ U_{i} $, in fact one should consider the rank of matrix $ K $ throughout the phase space. There may exist subregions of phase space where $ K $ is not full rank. In such cases no more the sets of constraints $\mathcal{S}_{i} $ and $ U_{i} $ are equivalent. In fact, one should use independent combinations of   $ U_{i} $ together with equations which determine the subregion where the rank of $ K $ is lowered. Note in general every multiplicative set of constraints (even in the form of matrices) should be broken in different branches of satisfaction. In other words, it is not allowed to use  the original constraints $\mathcal{S}_{i} $. To see an interesting case of this point see   Ref. \cite{Haji}. It is well known,  on the other hand, that \cite{zms2} given the constraints $ \varphi_{_{a}} $, one can redefine them as $ \varphi^{\prime}_{a}=M_{ab}\varphi_{b} $ provided that $ M_{ab} $ is nonsingular on the constraint surface. Otherwise, it is obvious that the constraint structure may change. For example if we multiply the second class constraints with some other second class constraints we would find first class constraints. } 

Another subtlety concerns the Eq. (\ref{bi13}) as an eigenvalue problem for the matrix $\frac{\partial(D^{k}_{\ j}n^{j})}{\partial n^{i}}$ with eigenvalue $\frac{-M}{N}$ and eigenvector $U_{k}$. However, since $U_{k}$ is a definite vector given in Eq. (\ref{bi15}) this exceptional case does not matter.
 
Therefore, we have 8 primary constraints  $P_{N},P_{M},P_{M^{i}},P_{n^{i}}$ and 8 secondary constraints $\mathcal{C},\mathcal{D},\mathcal{R}_{i},U_{i}$.  
Now we should consider the consistency of second level constraints. If we wrongly have not replaced the secondary constraints $\mathcal{S}_{i}  $ with $U_{i}$
, the matrix $ \partial^{2}V/\partial L^{a}\partial L^{b} $ in equation (\ref{m15}) would have rank 5, since terms  $ \partial^{2}V/\partial n^{i}\partial N$
and  $ \partial^{2}V/\partial n^{i}\partial M$ do not vanish (although  $ \partial^{2}V/\partial n^{i}\partial M^{i}$ vanish). This fact contradicts our expectation to have at least  4 null-vectors for  $ \partial^{2}V/\partial L^{a}\partial L^{b} $. However, considering the secondary constraint $ U_{i} $ make the  $ 8 \times 8 $ matrix $ \partial \tilde{A}_{a}/\partial L^{b} $ such that the first five columns and  the first five rows  vanish and only the elements $ \partial^{2}U_{i}/\partial n^{i} $  are non vanishing. 

Putting all together, the consistency equations for the second level constraints, i.e. Eq. (\ref{jj1}), reads 
\begin{equation}
\left(
\begin{array}{cr}
\{\mathcal{C}, H_{c}\}  \\
\{\mathcal{D}, H_{c}\}    \\
\{\mathcal{R}_{i}, H_{c}\}  \\
\{U_{i}, H_{c}\}  \\
\end{array}
\right)
+
\left(
\begin{array}{c|c}
\textbigcircle  &\textbigcircle    \\
 & \\
\hline
 &\\
\textbigcircle&   \partial U_{i}/\partial n^{j}  \\
\end{array}
\right)
\left(
\begin{array}{cr}
u    \\
v   \\
u^{i}  \\
v^{i} \\
\end{array}
\right)=0.
\end{equation}
As is seen, the null vectors of the matrix $ \partial \tilde{A}_{a}/\partial L^{b} $ give the third level constraints $ \{ \mathcal{C},H_{c}\} $, $ \{ \mathcal{D},H_{c}\} $ and $ \{ \mathcal{R}_{i},H_{c}\} $ respectively. The constraints $ \mathcal{R}_{i}$  have vanishing Poisson brackets with all the primary as well as secondary constraints, as calculated in full details in Refs. \cite{HR5, jklu}. Since the canonical Hamiltonian (\ref{bi5}) is composed of secondary constraints, the Poisson brackets $ \{ \mathcal{R}_{i},H_{c}\} $  also vanish. This shows that consistency of $ \mathcal{R}_{i}$ neither determines any of the Lagrange multipliers nor leads to any further constraint. 
Since $ \mathcal{R}_{i}$ is the sum of momentum constraints due to the individual Einstein-Hilbert actions of $ g_{\mu \nu} $ and $ f_{\mu\nu} $, we expect the set of 6 constraints $ P_{M^{i}} $ and $ \mathcal{R}_{i}$ to act as generators of the spacial diffeomorphisms. 

Putting aside the 6 first class constraints $P_{M^{i}}$ and $\mathcal{R}_{i}$, there remain constraints $P_{N},P_{M}$ and $ P_{n^{i}} $ as primary constraints and $ \mathcal{C},\mathcal{D} $ and $ U_{i} $ as secondary constraints. 
Since $ U_{i} $ are functions of $n_{i}$ such that $ \lvert \frac{\partial U_{i}}{\partial n_{k}}\rvert \neq 0$, the set of six constraints $ P_{n^{i}} $ and $ U_{i} $ are second class. Hence, consistency of $ U_{i} $'s leads to determination of Lagrange multipliers $v^{i}$'s in Eq. (\ref{k111}). These second class constraints should be imposed strongly on the system, in order to reach the reduced phase space. Hence, from now on, the momenta $P_{n^{i}} $ should be considered as zero and due to  $U_{i}=0 $, the variables $n_{i} $ would be determined in terms of the canonical variables $ g_{ij}, \pi^{ij}, f_{ij}$ and $p^{ij} $.

Now we should investigate the time evolution of $ \mathcal{C}$  and $\mathcal{D} $. Remember that the Poisson brackets of $ \mathcal{C}$  and $\mathcal{D} $ with $\mathcal{R}_{i}$ vanish since $\mathcal{R}_{i}$ are first class. Moreover, it is directly seen that  $\lbrace \mathcal{C}, p_{n^{j}} \rbrace= U_{i}\frac{\partial(D^{i}_{\ k}n^{k})}{\partial n^{j}}\approx 0 $ and $\lbrace \mathcal{D}, p_{n^{j}} \rbrace=U_{i}\approx 0$ which vanish weakly \cite{HR5}. It can also be shown that $ \lbrace \mathcal{C}(x),\mathcal{C}(y) \rbrace \approx 0 $ and $ \lbrace \mathcal{D}(x),\mathcal{D}(y) \rbrace \approx 0 $ \cite{HR5, jklu}. Hence, consistency of the constraints $ \mathcal{C}$  and $\mathcal{D} $ by using the canonical Hamiltonian (\ref{bi5}) gives the following third level constraints, 
\begin{eqnarray} &&
\lbrace \mathcal{C}(x),H_{c} \rbrace =\int d^{3}yM(y)\lbrace  \mathcal{C}(x),\mathcal{D}(y) \rbrace=M(x)\Gamma(x), \label{k112}
\end{eqnarray}
\begin{eqnarray} &&
\lbrace \mathcal{D}(x),H_{c}\rbrace =\int d^{3}y N(y)\lbrace \mathcal{C}(x),\mathcal{D}(y) \rbrace=N(x)\Gamma(x).\label{k113}
\end{eqnarray}
where 
\begin{eqnarray} &&
\Gamma\approx \left(\frac{m^{4}}{M^{2}_{g}}(g_{mn}\pi-2\pi_{mn})U^{mn}\right)+2m^{4}\sqrt{g}g_{ni}D^{i}_{\ k}n^{k}\triangledown_{m}U^{mn}\nonumber\\&& \hspace{5mm}+\left(\mathcal{R}_{j}^{(g)}D^{i}_{\ k}n^{k}-2m^{4}\sqrt{g}g_{ik}\bar{V}^{ki}\right)
\triangledown_{i}n^{j}\nonumber\\&&\hspace{5mm}+\sqrt{ g}\left(\triangledown_{i}(\mathcal{R}^{0(g)}/\sqrt{ g})+\triangledown_{i}(\mathcal{R}_{j}^{(g)}/\sqrt{ g})D^{j}_{\ k}n^{k}\right)n^{i}\nonumber\\&&\hspace{5mm}-\frac{m^{4}}{M^{2}_{f}}\frac{\sqrt{ g}}{\sqrt{ f}}\left(f_{mn}p-2p_{mn}\right)\bar{F}^{mn}, \label{gama}
\end{eqnarray}
in which
\begin{eqnarray} &&
\ U^{mn}=-\sqrt{x}g^{mn},\\&&
\bar{V}^{ki}=g^{kj}(-\frac{f_{jl}}{\sqrt{x}}((D^{-1})^{l}_{\ r}g^{ri})),\\&&
\bar{F}^{mn}=-\frac{(D^{-1})^{m}_{\ i}g^{ni}-n^{i}n^{m}D^{n}_{\ i}}{\sqrt{x}}.
\end{eqnarray}

Historically this point is the most crucial point in the investigation of bi-gravity and proving that it is ghost free. Obviously the Poisson bracket $ \{\mathcal{C}(x),\mathcal{D}(y) \} $ is nonzero which states both $ \mathcal{C}$ and $ \mathcal{D} $ are second class constraints. In Ref. \cite{HR5} which is the main reference of so many papers using HR model, it is argued that $\mathcal{D}$ is first class "since  we need it to be first class in order to generate diffeomorphism". Hence, the authors of \cite{HR5}, just "assume" that $ \Gamma\equiv \{\mathcal{C}(x),\mathcal{D}(y) \} $ is a new constraint (they denote it as $ \mathcal{C}_{2} $ ) which constitute a system of second class constraints together with the constraint $ \mathcal{C} $. 

Two important points arise here. First, there is no preference  between $ \mathcal{C} $ and $\mathcal{D}$. One could choose $ \mathcal{C} $ instead of   $ \mathcal{D} $ as a first class constraint which may generate guage transformations. In fact, it requires  complicated calculations to find which one of  $ \mathcal{C}$  or $ \mathcal{D} $, or a combination of them, is the generator of diffeomorphism. Second, with this logic one may consider $ \Omega\equiv \{\mathcal{C},\Gamma \}$ as a new constraint and claim that $ \mathcal{C} $ is also first class. This story may have no end. In fact, in the general context of constrained systems the Poisson bracket of second class constraints just act as non vanishing coefficients in determining the Lagrange multipliers\cite{BGP} and it is not reasonable to consider them as new constraints. New constraints at each level come out only as the Poisson brackets of the existing constraints with the canonical Hamiltonian. 

This point about the pioneer paper \cite{HR5} was also observed by Kluson in Ref.  \cite{jklu}. He investigated similar Hamiltonian analysis as we gave briefly in this section, up to the bottle neck of calculating $ \{\mathcal{C}(x),\mathcal{D}(y) \} $. He found that one is not able to obtain a new constraint out of consistency of constraints  $ \mathcal{C} $ and $\mathcal{D}$. However, in Ref. \cite{jklu} the following differential equations are derived  for consistency of  $ \mathcal{C} $ and $\mathcal{D}$  respectively.
\begin{eqnarray} &&
\mathcal{C}_{2}\equiv M(F-\partial_{i}V^{i})+(W^{i}-V^{i})\partial_{i}M\approx 0,
\nonumber\\ &&
\mathcal{D}_{2}\equiv N(F-\partial_{i}V^{i})+(W^{i}-V^{i})\partial_{i}N\approx 0, \label{diff}
\end{eqnarray} 
with some expressions for $ F$, $V^{i} $ and $W^{i}  $. Since the constraints  $ \mathcal{C} $ and $\mathcal{D}$  contain spacial derivatives of the canonical variables, it does not seem strange to obtain derivatives of the delta function in the Poisson bracket $ \{\mathcal{C}(x),\mathcal{D}(y) \} $ which lead to differential equations for $ M $ and $ N $ respectively.
 
 Eqs. (\ref{diff})  show the constraints $ (\mathcal{C}_{2} , \mathcal{D}_{2} )$  together with $ (P_{M}, P_{N}) $ and $ (\mathcal{C},\mathcal{D}) $ constitute a system of 6 second class constraints. In this way we have, at one hand, a system of  6 first class and 12 second class constraints leading to 16 phase space degrees of freedom which involves ghost. On the other hand, the gauge symmetry is restricted to spacial diffeomorphism generated by $ P_{M_{i}} $
and $\mathcal{R}_{i}  $.  This objection concerning the existence of ghost in HR model remained unanswered 
for almost four years. 

In our study of  this problem, for time evaluation of the constraints  $ \mathcal{C}$ and $ \mathcal{D} $,  we observed that both the constraints $ \mathcal{C}_{2} $ and $ \mathcal{D}_{2} $ in Kluson's analysis are of the same structure as   $ \mathcal{C}_{2}(x)=\int d^{3}y \Gamma(x,y) M  $ and  $ \mathcal{D}_{2}(x)= \int d^{3}y\Gamma(x,y) N $ where $ \Gamma(x,y)\equiv \{\mathcal{C}(x),\mathcal{D}(y) \} $ may contain derivatives of delta function. Apart from dependence of $ \Gamma(x,y) $ on derivative of the delta function, one may consider the consistency equations $ \dot{\mathcal{C}}=\Gamma M=0 $ and $ \dot{\mathcal{D}}=\Gamma N=0 $ as a bifurcation problem. In other words, one may consider these equations as something  to determine  $ M $ and $ N $ or they can be satisfied just by one condition $ \Gamma =0 $. We are mostly familiar with cases where $ \Gamma(x,y) $ is proportional to $ \delta(x-y) $. However, in a formal way, one may also consider the case where $ \Gamma $ also contains $ \partial_{i}\delta(x-y) $. 

During the weeks we were preparing this article a new paper by F. Hassan and A. Lundkvist \cite{Hassan18} was published which shows the correct expressions of $ \mathcal{C}_{2} $ and $ \mathcal{D}_{2} $ do not contain derivatives of  $ M $ and $ N $. Our calculations are also in agreement with this results. In other words, the Poisson bracket $ \{\mathcal{C},\mathcal{D} \} $ does not contain derivatives of delta function, at all. Hence, the consistency conditions of  $ \mathcal{C}_{2} $ and $ \mathcal{D}_{2} $ do not give equations (\ref{diff}) , but they give $  \Gamma N\approx0$ and $ \Gamma M\approx0 $ where $ \Gamma(x) $
is as given in Eq. (\ref{gama}), in agreement with the result of \cite{Hassan18}.

We emphasize again that the system of equations 
 $  \Gamma N\approx0$ and $ \Gamma M\approx0 $ are, in fact, a real bifurcation problem, where you need to make a choice to proceed with the problem. Here we have two choices: 
 
 i) Every where in phase space where $ \Gamma $ does not essentially vanish, we should impose $ M\approx N\approx0 $, which is more or less similar to the result of Ref. \cite{jklu} discussed above, i.e. 16 degrees of freedom containing ghost and lack of complete four parameter diffeomorphism of space-time. The worst result of the choice $ \Gamma \neq 0 $, $ N \approx M \approx 0 $ is emerging singular metrics which are not acceptable physically. However, note that the dynamics of theory, by itself, does not discard this possibility. This is our physical preference to put this choice away, which is imposed from outside of the dynamical investigation of the model.
 
 ii)  If we restrict ourselves to the subregion $ \Gamma \approx 0 $ of the phase space, we would have no restriction on the lapses $ N $ and $ M $ up to this point, i.e. they remain arbitrary so far. However, we may encounter some restrictions on the lapse functions (not here but) in the subsequent levels of canonical investigation of the theory, as we will see. 
 
 However, as we mentioned before, $ \Gamma $ is not an ordinary constraint which comes out from the Poisson brackets of the existing constraints with the canonical Hamiltonian, as is the case, in Dirac approach, for every constraint system. In other words, $ \Gamma=0 $ is not a natural consequence of the dynamics of the system; it is just a kind of constraint or restriction on the canonical variables which you impose, in order to escape unwanted results  $ M= 0 $ and $ N = 0 $. 
 
 Hence, in our opinion, it needs special care to see what naturally emerges from the dynamics of the theory and what we "assume" in order to have a consistent theory. In fact, constraints such as $ \Gamma $ should be viewed as a new kind of constraints, which are different from primary constraints (which emerge due to definition of momenta) and secondary constraints (which emerge from the Poisson brackets of the constraints with the canonical Hamiltonian). This kind of constraints which we denote them as "new kind" are also familiar to us in the canonical analysis of Chern-Simons like theories in 3 dimensions \cite{Haji}. \footnote{The necessity of additional constraints has been observed previously in canonical analysis of Chern-Simons like theories in Ref. \cite{x1}-\cite{x3}. However, their special character and distinguishing them from normal Dirac constraints were not recognized before.}
 
 Assume, any how, that we have accepted $ \Gamma $ as a new constraint. It is obvious that the system should not exit from the surface $ \Gamma=0 $ during the time evaluation. Hence, the consistency condition  $ \dot{\Gamma}=0 $ should be imposed further. This gives the fourth level constraint 
\begin{eqnarray} &&
\Omega(x)\equiv \int d^{3}y \{\Gamma(x), H_{c}(y)\}_{*}=E(x)M(x)+F(x)N(x)
\end{eqnarray}
 where 
\begin{eqnarray} &&
F(x)N(x)=\int d^{3}y N(y)\{\Gamma(x), \mathcal{C}(y)\}_{*}\\&&
E(x)M(x)=\int d^{3}y M(y)\{\Gamma(x), \mathcal{D}(y)\}_{*}
\end{eqnarray}
The symbol $\lbrace \ ,\ \rbrace_{*}$  means the Dirac bracket \cite{dms3} which implies strongly imposing the constraints $ p_{n^{i}} $ and $ U_{i} $ (see Eq. (\ref{bi15}) ).
The constraint $ \Omega(x) $ contains the lapse functions $ M $ and $ N $. So one combination of the Lagrange multipliers $ u $ and $v$ in the total Hamiltonian would be determined from 
consistency of $ \Omega $, i.e. 
\begin{eqnarray} &&
\int d^{3}z \lbrace \Omega(x),\mathcal{H}_{T} (z)\rbrace_{*} =\nonumber\\&&
\int d^{3}z \lbrace \Omega(x),(\mathcal{H}_{c}+uP_{N}+vP_{M}) (z)\rbrace_{*} \approx 0. \label{op1}
\end{eqnarray}

The good news is that this is the end of the consistency process and one combination of $ u $ and $ v $ remain undetermined. In addition to the Lagrange multipliers $u_{i}$ in Eq. (\ref{k111}) we have, in this way, altogether 4 arbitrary gauge fields corresponding to diffeomorphism parameters. One may manage the whole structure of the problem in a more clear form if one changes the lapse variables to $ \bar{N}, M $ such that 
 \begin{eqnarray} &&
H_{c}=\bar{N}\mathcal{C}+M\mathcal{D}^{\prime}+M^{i}\mathcal{R}_{i}, \label{hc}
 \end{eqnarray}
 where
 \begin{eqnarray} &&
\bar{N}=N+\frac{E}{F}M\\&&
\mathcal{D}^{\prime}=\mathcal{D}-\frac{E}{F}\mathcal{C}
 \end{eqnarray}
In this system consistency of $ \mathcal{D}^{\prime} $ is satisfied identically on the surface $ \Gamma=0 $. Meanwhile, consistency of $ \Gamma $ gives $ \Omega=\bar{N}F $ and finally consistency of $ \Omega $ determine the Lagrange multiplier of the primary constraint $ P_{\bar{N}} $ in the total Hamiltonian. The interesting point is that at the final stage the problem bifurcates once more. In other words, we could restrict ourself on the surface $ F=0 $. However, we do not do this, since it makes our change of variables in Eq. (\ref{hc}) singular. Hence, our analysis is valid where $ F\neq0 $.

\section{Bi-gravity without square root}
To see better the general construction of section (2), let us consider a general model as  
\begin{equation}
S=\int d^{4}x \left( M^{2}_{g}  \sqrt{-^{(4)}g} 
R(g)+M^{2}_{f} \sqrt{-^{(4)}f} 
R(f)+2m^{4}  (^{(4)}g\ ^{(4)}f)^{1/4} V(\mathcal{Z}_{1},...,\mathcal{Z}_{4})\right), \label{m111}
\end{equation}
where $\mathcal{Z}_{n}=Tr[(g^{-1}f)^{n}] $.  Comparing to HR bi-gravity, this category concerns $\mathcal{Z}^{\mu}_{\ \nu}=g^{\mu \lambda}f_{\lambda\nu} $ instead of $\sqrt{\mathcal{Z}^{\mu}_{\ \nu}} $ . There is also a slight difference in coefficient of the interaction term where $ \sqrt{-^{(4)}g} $  is replaced by $ [^{(4)}g ^{(4)}f]^{1/4}$ which is more symmetric with respect to the $ g$ and $f $ metrics. 
For this case, it is convenient to use the following variables\cite{bul}
\begin{eqnarray}&&
\bar{N}= \sqrt{NM} ,\hspace{3mm}  n=\sqrt{\dfrac{N}{M}}, \hspace{3mm} \bar{N^{i}}=\dfrac{1}{2}(N^{i}+M^{i}),\hspace{3mm} n^{i}=\dfrac{N^{i}-M^{i}}{\sqrt{NM}}. \label{kogan1}
\end{eqnarray}
Considering Eqs. (\ref{g1}) and (\ref{g2}) together with Eq. (\ref{kogan1}), one can show directly
\begin{eqnarray}&&
\mathcal{Z}_{1}=\mathcal{Z}_{\ \mu}^{ \mu}=a+a_{i}^{i},\label{u11}\\&&
\mathcal{Z}_{2}=\mathcal{Z}_{\mu}^{\ \nu}\mathcal{Z}_{\ \nu}^{\mu}=a^{2}+v_{i}w^{i}+a_{j}^{i}a_{i}^{i} \label{u12}\\&&
\mathcal{Z}_{3}=\mathcal{Z}_{\ \mu}^{\rho}\mathcal{Z}_{\nu}^{\ \mu}\mathcal{Z}^{\nu}_{\ \rho}=a^{3}+3v^{i}w_{i}a+3v_{i}a_{j}^{i}w^{j}+a_{j}^{i}a_{k}^{j}a_{i}^{k} \label{u13}\\&&\mathcal{Z}_{4}=\mathcal{Z}_{\ \sigma}^{\rho}\mathcal{Z}_{\rho}^{\ \sigma}\mathcal{Z}_{\nu}^{\ \mu}\mathcal{Z}^{\nu}_{\ \mu}
\nonumber\\&& \hspace{6mm}=a^{4}+4a^{2}v_{i}w^{i}+2(v_{i}w^{i})^{2}+4av_{i}a^{i}_{j}w^{j}+4v_{i}a_{j}^{i}a_{k}^{j}w^{k}+a_{j}^{i}a_{k}^{j}a_{l}^{k}a_{i}^{l}, \label{u1}
\end{eqnarray}
where
\begin{eqnarray}&&
v_{i}=\frac{f_{ij}n^{j}}{n^{2}},\\&&
a=\frac{1}{n^{4}}-\frac{n^{i}f_{ij}n^{j}}{2n^{2}},\label{u21}\\&&
a^{i}_{j}=g^{ik}f_{kj}-\frac{n^{i}f_{jk}n^{k}}{2n^{2}},\label{u31}\\&&
w^{i}=n^{i}\frac{n^{m}f_{mn}n^{n}}{4n^{2}}-\frac{n^{i}}{2n^{4}}-\frac{1}{2}g^{im}f_{mk}n^{k}.
\end{eqnarray}
These relations show that the interaction potential  $ V(\mathcal{Z}_{1},...,\mathcal{Z}_{4}) $ is fortunately independent of $\bar{N} $ and $\bar{N}^{i} $. This enables us to linearize the action with respect to $\bar{N} $ and $\bar{N}^{i} $.
In Ref. \cite{Kluson} it  is argued that the characteristic equation of the matrix $\mathcal{Z}_{\ \nu}^{\mu} $ is the same as $A^{\mu}_{\ \nu}=\mathcal{Z}_{\ \nu}^{\mu}|_{\bar{N}=1,\bar{N}^{i}=0} $. Hence, it is deduced that, in principle, there exists a similarity transformation which brings $ \mathcal{Z}_{\ \nu}^{\mu} $ to $ A_{\ \nu}^{\mu} $. However, besides to direct calculations of $\mathcal{Z}_{1}$ to $\mathcal{Z}_{4}  $ in Eq. (\ref{u11}) to (\ref{u1}), we can simply argue that since $ Tr(\mathcal{Z}^{\mu}_{\ \nu})^{n} $ is gauge invariant; one can in fact calculate the corresponding quantities $ \mathcal{Z}_{n} $ in a special gauge where $ \bar{N}=1 $ and $\bar{N}^{i}=0  $, which gives the same results. 

Including the well-known result for the Einestain-Hilbert parts of the action (\ref{m111}), the Lagrangian density reads as Eq. (\ref{m12}) where 
\begin{equation}
H_{c}=\int d^{3}x (\bar{N}\mathcal{R}+\bar{N}^{i}\mathcal{R}_{i}),\label{hc2}
\end{equation}
in which 
\begin{equation}
\mathcal{R}=n\mathcal{R}_{0}^{(g)}+\frac{1}{n}\mathcal{R}_{0}^{(f)}+\frac{1}{2}n^{i}\mathcal{R}_{i}^{(g)}-\frac{1}{2}n^{i}\mathcal{R}_{i}^{(f)}+2m^{4} (gf)^{1/4}V(\mathcal{Z}^{\mu}_{\ \nu}),\label{k1}
\end{equation}
\begin{equation}
\mathcal{R}_{i}=\mathcal{R}_{i}^{(g)}+\mathcal{R}_{i}^{(f)}.
\end{equation}
As usual, the momenta conjugate to the lapse-shift variables $\bar{N},\bar{N}^{i},n$ and $n^{i}$  are primary constraints, i.e.
\begin{equation}
P_{\bar{N}}\approx 0,  P_{i}\approx 0,p_{n}\approx 0, p_{i}\approx 0 .
\end{equation}
Hence, the total Hamiltonian is as follows
\begin{eqnarray}
H_{T}=\int d^{3}x (\bar{N}\mathcal{R}+\bar{N}^{i}\mathcal{R}_{i}+ uP_{\bar{N}}+u_{i}P_{i}+v^{i}p_{i}+vp_{n}). \label{l1}
\end{eqnarray}
The time evaluation of the primary constraints gives
\begin{eqnarray} &&
\lbrace P_{\bar{N}},H_{T} \rbrace =-\mathcal{R}\approx 0,\label{bi110}\\ &&
\lbrace P_{i},H_{T}\rbrace = -\mathcal{R}_{i} \approx 0,\label{bi111}\\ &&
\lbrace p_{n} ,H_{T} \rbrace =\bar{N}(-\mathcal{R}_{0}^{(g)}+\frac{1}{n^{2}}\mathcal{R}_{0}^{(f)}-2m^{4} (gf)^{1/4}\frac{\delta V }{\delta n}) \equiv \bar{N}\zeta,\label{bi112}\\ &&
\lbrace p_{i},H_{T}\rbrace =\bar{N}(
-\frac{1}{2}\mathcal{R}_{i}^{(g)}+\frac{1}{2}\mathcal{R}_{i}^{(f)}-2m^{4} (gf)^{1/4}\frac{\delta V }{\delta n^{i}})\equiv \bar{N}\zeta_{i}.\label{bi113}
\end{eqnarray}
Comparing to our general formalism of section (2), the secondary constraints $ \mathcal{R}_{a} $ of Eq. (\ref{m14}) are $ \mathcal{R} $, $ \mathcal{R}_{i} $, $ \tilde{\zeta}\equiv  \bar{N}\zeta$ and $ \tilde{\zeta}^{i}\equiv  \bar{N}\zeta^{i}$ respectively. The constraints $ \mathcal{R}_{i} $ are mainly composed of the Einestain-Hilbert parts $ \mathcal{R}_{i}^{(g)} $ and $ \mathcal{R}_{i}^{(f)} $ and commute with each other. The constraint $ \mathcal{R} $ (see Eq. (\ref{k1})) is the most important part of the theory which includes the interaction term. Straightforward calculations given in Ref. \cite{Kluson} show $\lbrace \mathcal{R}(x),\mathcal{R}(y)\rbrace \approx 0 $ as well as $\lbrace \mathcal{R}(x),\mathcal{R}_{i}(y)\rbrace \approx 0 $. Taking a look on the secondary constraints $ \tilde{\zeta}$ and $\tilde{\zeta}_{i} $ shows that we have a bifurcation problem here. 

We are in general free to assume different cases  $\bar{N}=0 $ and $ \bar{N}\neq 0 $. The first choice, leads to a degenerate metric which is not physical. Hence, the simplified constraints $ \zeta $ and $ \zeta^{i} $ are resulted from the physical assumption $\bar{N}\neq 0  $. Let us note briefly that, in spite of the approach of Ref. \cite{Kluson}, it is not needed to add the secondary constraints to the total Hamiltonian. In fact, theoretically as shown in \cite{zms2,BGP}, the total Hamiltonian as the generator of time evaluation should only include the primary constraints.\footnote{Working with the extended Hamiltonian,  which includes all  constraints may sometimes simplify the problem, but not for the case at hand. However, the extended Hamiltonian turns out to give the correct time evaluation for the gauge invariant quantities.} Adding the secondary constraints to the total Hamiltonian, however, makes us to calculate some unnecessary Poisson brackets. 

Now we need to consider the consistency of secondary constraints, by using the total Hamiltonian (\ref{l1}). This should give us equations similar to Eq. (\ref{m15}) for $u_{a}$'s as unknowns. Since $R_i$'s  include non of the laps and shift functions they would commute with all of the primary constraints, as well as the canonical Hamiltonian.  The constraint $R$, however, do depend on $n$ and $n_i$ (see Eq. \ref{k1}). It is easy to see that $\{ R, p_\mu \}=\partial R/\partial n_\mu = \xi_\mu \approx 0$. Hence, consistency of the constraints $R$ and $R_i$ gives no new constraint and determines non of the Lagrange multipliers. Therefore, the only non-trivial part of  Eq. (\ref{m15}) comes from the consistency of the constraints $ \zeta $ and $\zeta_{i}  $. In this way the consistency conditions of secondary constraints can be given by the following matrix 
\begin{equation}
\left(
\begin{array}{cr}
0  \\
0    \\
\{\zeta , H_{c}\}  \\
\{\zeta_{i}, H_{c}\}  \\
\end{array}
\right)
+
\left(
\begin{array}{c|c}
\textbigcircle & \textbigcircle    \\
&  \\
\hline
& \\
\textbigcircle &  \vartriangle_{\mu\nu}  \\
\end{array}
\right)
\left(
\begin{array}{cr}
u    \\
u^{i}  \\
v \\
v^{i} \\
\end{array}
\right)=0.
\end{equation}
where
\begin{eqnarray} &&
\vartriangle_{\mu\nu}\equiv \lbrace \zeta_{\mu},p_{\nu} \rbrace=\frac{\partial^{2} \tilde{V}}{\partial n^{\mu}\partial n^{\nu}}, \label{ty}
\end{eqnarray}
in which 
\begin{eqnarray} &&
 \tilde{V}=\frac{1}{n}\mathcal{R}_{0}^{(f)}+2m^{4} (gf)^{1/4}V. \label{1ty}
\end{eqnarray}

As expected, the matrix of the coefficients of $ u $'s and $ v $'s have four null-vector, which do not lead to any new constraint. Hence, the four variables $ u $ and $ u^{i} $ remain undetermined through dynamical investigation of the theory. 
The only nontrivial part of the consistency procedure of the secondary constraints then reads 
\begin{equation}
\{\zeta_{\mu} , H_{c}\}-\vartriangle_{\mu\nu}v^{\nu}=0,\label{5k}
\end{equation}

If $ \det(\vartriangle_{\mu\nu})\neq 0 $,  the constraints  $ p_{n},\zeta  $ and $ p_{i},\zeta_{i} $ are second class.
Hence, we have 8 first class and 8 second class constraints which gives 16 dynamical phase space variables, (see Eq. (\ref{hy})). This involves a scalar ghost.  If $ \det (\vartriangle_{\mu\nu})=0 $, 
we should have at least one null vector for the matrix $ \vartriangle_{\mu\nu} $, denoted by $ \lambda^{\mu} $. Multiplying Eq. (\ref{5k}) by $ \lambda^{\mu} $, we find the new constraint $ \lambda^{\mu}\{\zeta_{\mu} , H_{c}\}=0 $. 
Since $ \{\zeta_{\mu} , \mathcal{R}_{i}\}=0, $ from  Eq. (\ref{hc2}) the new constraint reads 
\begin{eqnarray} &&
\lambda^{\mu}\lbrace \zeta_{\mu}(x),H_{c}(y) \rbrace =\int d^{3}y \bar{N}(y)\left(	\delta(x-y)\mathcal{F}(x)+\mathcal{W}^{i}(x)\partial_{x^{i}}\delta(x-y)\right)\approx0, \label{mei210}
\end{eqnarray}
for some functions $\mathcal{F}(x)  $ and  $ \mathcal{W}^{i}(x) $. If $ \mathcal{W}^{i}(x)\neq 0 $, the equation (\ref{mei210}) gives a differential equation for the lapse function $ \bar{N} $. However, from the requirement of diffeomorphism, we need  $ \bar{N} $ to be an arbitrary field, while a differential equation restricts our arbitrariness only to it's initial condition. Ref. \cite{Kluson} deduces from this point that the case $ \det (\vartriangle_{\mu\nu})=0 $, should not happen; hence all models of the form of Eq. (\ref{m111}), including HR bi-gravity contain ghost mode. However, as pointed out in a footnote in the same reference, there is the possibility of vanishing $ \mathcal{W}^{i}(x) $, which changes the constraint (\ref{mei210}) to bifurcation form $ \bar{N}(x)\mathcal{F}(x) $. Again, we use the physical condition $ \bar{N}\neq 0 $ to consider $ \mathcal{F}(x) $ as a new constraint. Consistency of $ \mathcal{F}(x) $ may also lead to a differential equation for $ \bar{N}$. If we are enough lucky, the coefficient of the derivative of delta function in this new equation may also vanish. Under these circumstances, we would have two more second class constraints which cancel the ghost. Although, it seems too improbable, however, the analysis of HR gravity for the more complicated potential (involving the square root of $ g^{-1}f $) shows that it may be possible for a specially designed interaction potential to reach the desired two more constraints needed to omit the ghost. 

 We want here to be bold enough to give a new suggestion. Consider the differential equation (\ref{mei210}) for $ \bar{N}$ as an integral equation 
\begin{eqnarray} &&
\int d^{3}y \Upsilon(x,y)\bar{N}(y)\approx0, \label{meki210}
\end{eqnarray}
This can also be considered as a bifurcation problem for the two factors $ \tilde{\Upsilon}(x,y) $ and $ \bar{N}$. Hence, implying $ \bar{N} \neq 0$ may lead us to consider the new constraint $ \Upsilon(x,y)\equiv\delta(x-y)\mathcal{F}(x)+\mathcal{W}^{i}(x)\partial_{x^{i}}\delta(x-y)\approx0 $.
This kind of constraint is deviated slightly from being a local constraint; so we denote it as a "semi local constraint". Consistency of $ \Upsilon $ may give us again a semi local constraint. We think these new constraints are still strong enough to omit the ghost degree of freedom. However, further details requires to consider a given model of the form given in Eq. (\ref{m111}). Here, we just suggested the idea.

To see the above arguments better, consider the concrete example in which the interaction term is $ V=\mathcal{Z}_{1} $ as given in Eq. (\ref{u11}), which is also analyzed in Ref. \cite{Klu3a}. Using Eqs. (\ref{u21}) and (\ref{u31}) we have 
\begin{equation}
V=\frac{1}{n^4}-\frac{n^if_{ij}n^j}{n^2}+g^{ij}f_{ij}.\label{111}
\end{equation}
For this particular interaction the constraints $\zeta$ and $ \zeta_{i} $ and the matrix $  \vartriangle_{\mu\nu}$ read
\begin{eqnarray}&&
\zeta\equiv -\mathcal{R}_{0}^{(g)}+\frac{1}{n^{2}}\mathcal{R}_{0}^{(f)}-2m^4(gf)^{1/4}\left( \frac{-4}{n^{5}}+\frac{2n^{i}f_{ij}n^{j}}{n^{3}}\right),\\&&
\zeta_{i} \equiv\frac{-1}{2}\mathcal{R}_{i}^{(g)}+\frac{1}{2}\mathcal{R}_{i}^{(f)}+2m^{4}(gf)^{1/4}(\frac{2n^{i}f_{ij}}{n^{2}}),
\end{eqnarray}

\begin{eqnarray}&&
\vartriangle_{\mu\nu}=\left(
\begin{array}{cr}
 -2\mathcal{R}_{0}^{(f)}/n^3 +\alpha(-20/n^6 +6n^i f_{ij}n^j/n^4)  & -4\alpha n^if_{ij}/n^3 \\
   &   \\
-4\alpha n^i f_{ij}/n^3 & 2\alpha f_{ij}/n^2  \\
\end{array}
\right), \label{e1}
\end{eqnarray}
where $ \alpha \equiv (2m^4(g)^{1/4}(f)^{1/4}) $.
To find the probable null-vector of the matrix $ \vartriangle_{\mu\nu} $ first consider the last three columns which are proportional to $ \left(
\begin{array}{cr}
2n^{i}f_{ij}   \\
nf_{ij} \\
\end{array}
\right)  $. Since $f_{ij}$ is considered to be non-singular, each null-vector $ \lambda^{\mu} $ of $  \vartriangle_{\mu\nu}$ should necessarily be of the form $ (n,2n_i) $. However, such a vector obviously have not vanishing product with the first column. Moreover, direct calculation shows 
$  \vartriangle_{ \mu\nu}$ in Eq. (\ref{e1}) is nonsingular. This analysis indicates a bi gravity theory with interaction $ V=\mathcal{Z}_{1} $ consists ghost. However, one may consider more complicated interactions including $ \mathcal{Z}_{2} $, $ \mathcal{Z}_{3} $ and $ \mathcal{Z}_{4} $ in Eqs. (\ref{u11}-\ref{u1}). Theoretically it is not impossible to have an interaction for which the matrix $\vartriangle_{\mu\nu}  $ is singular, and subsequent conditions for a ghost-free theory of bi gravity are satisfied. However, finding such a model seems to be a second realization of the old dream of having ghost free bi gravity theory (after HR model). 

\section{Conclusions }
We performed the Hamiltonian analysis of  four dimensional bi-gravity theories in the context of ADM formalism. First, we worked in the framework of the original lapse and shift variables. In order to generate the gauge symmetry, i.e. the diffeomorphism, in the 40 dimensional phase space, we need to have 8 first class constraints in the first and second level of consistency of constraints. Hence, the matrix of the second derivatives of the interaction term with respect to lapse and shift variables should at least have 4 null-vectors. However, if we demand omitting the ghost, we need one more null-vector. 

This structure is preserved in every reparametrization of the lapse and shift functions. In fact, the main work done in reference \cite{HR5} is to find a suitable change of variables, so as to show that for HR bi-gravity the $8 \times 8$ matrix of second derivatives of the potential term has rank three with respect to the new variables. Note, however, that nobody has claimed this characteristics is exclusive for HR bi-gravity. Although difficult, it is not impossible for future model builders to introduce new models with the same property.

Suppose that the first condition is fulfilled and we have five constraints at the second level which are not second class so far. If consistency of these constraints gives no third level constraint, then we would have 6 second class and 10 first class constraints which corresponds to a ghost free model with 14 degrees of freedom. However, such a model would have one more gauge symmetry besides diffeomorphism. Theoretically  it does not seem impossible to have a model of this kind, but there is no known model of this category.

It is more or less known  that 6 first class constraints, which generate the spacial diffeomorphism, can easily be found in every covariant model of bi-gravity. Hence, the only way to have a ghost free theory of bi-gravity is finding two more second class constraint after the second level. Unfortunately, our demand is not satisfied in a straightforward manner. It seems that we usually find equations to determine the lapse functions due to consistency of the remaining constraints of second level. In other words, by no means one can find ordinary constraints which do not depend on the lapse functions in this procedure.  

Our important observation in this paper is that at this stage we have in fact a bifurcation problem. The theory, as it stands, may have dynamical sectors in which the lapse functions are constrained. This is in contradiction with our physical expectation that lapse function should act as part of gauge parameters in diffeomorphism. 

On the other hand, if we restrict ourselves to a limited subregion of the phase space described by additional constraints, the consistency condition of the remaining constraints may have different solution. In other words,  if we  assume that in the physical sector of the theory the lapse functions should not vanish (or determined severely) then the only consistent subregion is achieved by imposing additional constraint. As we found, this constraint in HR bi-gravity gives under consistency a fourth level constraint, whose consistency determines a special combination of lapse functions. 

We argued that even in case where the consistency condition of remaining second level constraint leads to differential equations, the bifurcation characteristics of the problem remains unchanged. In such cases we introduced the notion of semi-local constraints which contain some limited number of derivatives of delta function. 

The interesting point is the original model at the bifurcation point may go through the branch which fixes the lapse functions. If so, the theory has not advantage of the full capacity of four dimensional diffeomorphism; i.e. the gauge symmetry is limited to spacial diffeomorphism. This shows in the Hamiltonian framework we have additional situations which may not occur in Lagrangian formulation. 

However, through the physical branch, in addition to two second class constraints needed for omitting the ghost, we also have found two more first class constraints needed to generate the full four dimensional diffeomorphism. Unfortunately, this analysis just relies on counting the number of first class constraints. A difficult problem concerns how the variations of dynamical variables due to diffeomorphism is generated by these first class constraints. This may be the issue of our future works. 

As mentioned in the introduction, the bi-gravity models maybe employed in describing the observations concerning Neutron star merger GW170817 event. As stated in \cite{YA} this event puts constraints on the physical parameters of the bi-gravity coupled to matter. However, our investigation in this paper concerns only pure gravity and the problem of ghost. It is obvious that having a consistent theory of bi-gravity at hand, we are able to adjust its coupling to matter so as to fulfill the constraints imposed by the observations. 

\vspace{8mm}

{\bf{Acknowledgements:}} The authors would like to thank Claudia de Rham for helpful discussions. Z.M. thanks IPM for hospitality during the progress of this work.
\vskip .3cm

\vspace{8mm}
\appendix
\numberwithin{equation}{section}

\section{Some details of HR bi-gravity} \label{sec:rev}

\textbf{A)} \hspace{3mm} elementary symmetric polynomials $e_{n}(\Bbbk)$  as follows 
\begin{eqnarray}&&
e_{0}(\Bbbk)=1,\label{a2}
\nonumber\\ &&
e_{1}(\Bbbk)=[\Bbbk],\label{a3}
\nonumber\\ &&
e_{2}(\Bbbk)=\frac{1}{2}([\Bbbk]^{2}-[\Bbbk^{2}]),\label{a4}
\nonumber\\ &&
e_{3}(\Bbbk)=\frac{1}{6}([\Bbbk]^{3}-3[\Bbbk][\Bbbk^{2}]+2[\Bbbk^{3}]),\label{a5}
\nonumber\\ &&
e_{4}(\Bbbk)=\frac{1}{24} ( [\Bbbk]^{4}-6[\Bbbk]^{2}[\Bbbk^{2}] + 
3[\Bbbk^{2}]^{2}+8[\Bbbk][\Bbbk^{3}]-6[\Bbbk^{4}]),\label{a6}
\nonumber\\ &&
e_{i}(\Bbbk)=0,\ \ i > 4,\label{a7}
\end{eqnarray}
where $[\Bbbk] \equiv Tr(\Bbbk)$ and so on.

\textbf{B)} \hspace{3mm}   $D^{i}_{\ j}$ should be considered as
\begin{equation}
D^{i}_{\ j}=\sqrt{g^{id}f_{dm}W^{m}_{n}}(W^{-1})^{n}_{j},\hspace{3mm}  W^{l}_{\ j}=[1-n^{k}f_{km}n^{m}]\delta^{l}_{j}+n^{l}f_{mj}n^{m}.\label{a19}
\end{equation}

\textbf{C)} \hspace{3mm} The fundamental Poisson brackets to be used in  the canonical analysis are as follows
 \begin{eqnarray} &&
 \lbrace N(x),  P_{N}(y)\rbrace=\delta(x-y), \nonumber\\ &&
 \lbrace n^{i}, P_{n^{j}}(y)\rbrace=\delta^{i}_{\ j}  \delta(x-y), \nonumber\\ &&
 \lbrace g_{ij}(x),\pi^{kl}(y)\rbrace=1/2(\delta^{k}_{\ i}\delta^{l}_{j}+\delta^{l}_{\ i}\delta^{k}_{\ j})\delta(x-y),\nonumber\\ &&
 \lbrace f_{ij}(x),p^{kl}(y)\rbrace=1/2(\delta^{k}_{\ i}\delta^{l}_{\ j}+\delta^{l}_{\ i}\delta^{k}_{\ j})\delta(x-y),\nonumber\\ &&
 \lbrace M(x),  P_{M}(y)\rbrace=\delta(x-y), \nonumber\\ &&
 \lbrace M_{i}(x),  P_{M^{j}}(y)\rbrace=\delta^{i}_{\ j}\delta(x-y).\label{bi8}
 \end{eqnarray}

 \end{document}